\documentclass[aps,pra,twocolumn]{revtex4-1} 
\usepackage{verbatim}
\usepackage{dsfont} 
\usepackage{amsmath}
\usepackage{amsthm} 
\usepackage{enumerate}
\usepackage{graphicx}
\usepackage{mathtools}
\usepackage{bm} 
\usepackage{amssymb} 
\usepackage{tcolorbox}
\usepackage{hyperref} 
\usepackage{appendix} 
\usepackage[capitalize,nameinlink]{cleveref}
\hypersetup{
    colorlinks=true,
    linkcolor=blue,
    citecolor=blue,
    unicode=true,          
}

\usepackage{blindtext}

\newtheorem{result}{Result}
\newtheorem{lemma}{Lemma}

\newcommand{\ket}[1]{\left| #1 \right\rangle}
\newcommand{\bra}[1]{\left\langle #1 \right|}
\newcommand{\braket}[1]{\left\langle #1 \right\rangle}
\newcommand{\ketbra}[2]{\left|#1 \rangle \langle #2 \right|}

\newcommand{\ot}{\otimes}
\DeclareMathOperator{\tr}{Tr}

\newcommand{\ftel}{\mathbb{F}_{\small{\rm T}}}
\newcommand{\fbn}{\mathbb{F}_{\small{\rm BN}}}

\usepackage{calc} 
\usepackage{accents}

\usepackage{diagbox}
\usepackage{textpos}

\begin{document}
\title{All incompatible sets of measurements can generate Buscemi nonlocality}

\author{Andr\'es F. Ducuara$^{1,2,3}$} 
\email[]{andres.ducuara@yukawa.kyoto-u.ac.jp}

\author{Patryk Lipka-Bartosik$^{4, 5}$}
\email[]{patryk.lipka-bartosik@bristol.ac.uk}

\author{Cristian E. Susa$^{6}$}

\author{Paul Skrzypczyk$^{7}$}

\affiliation{$^{1}$Nanyang Quantum Hub, School of Physical and Mathematical Sciences, Nanyang Technological University, 637371, Singapore
\looseness=-1}

\affiliation{$^{2}$Yukawa Institute for Theoretical Physics, Kyoto University, Kitashirakawa Oiwakecho, Sakyo-ku, Kyoto 606-8502, Japan
\looseness=-1}

\affiliation{$^{3}$Center for Gravitational Physics and Quantum Information, Yukawa Institute for Theoretical Physics, Kyoto University
\looseness=-1} 

\affiliation{$^{4}$Center for Theoretical Physics, Polish Academy of Sciences, Warsaw, Poland
\looseness=-1} 

\affiliation{$^{5}$ Institute of Theoretical Physics, Jagiellonian University, 30-348 Kraków, Poland}

\address{$^{6}$Department of Physics and Electronics, University of C\'ordoba, 230002 Monter\'ia, Colombia\looseness=-1}

\affiliation{$^{7}$H.H. Wills Physics Laboratory, University of Bristol, Tyndall Avenue, Bristol, BS8 1TL, United Kingdom 
\looseness=-1}

\date{\today}

\begin{abstract}
    The presence of Bell-nonlocality in the correlations arising from measuring spatially-separated systems guarantees that the sets of measurements used are necessarily incompatible. Not all sets of incompatible measurements can however lead to Bell-nonlocality, as there exist incompatible sets of measurements which can only produce local correlations. In this work we prove that all sets of incompatible measurements are nevertheless able to generate nonlocality in an extended Bell scenario where quantum, instead of classical, measurement inputs are considered. In particular, this holds true for all incompatible-local sets of measurements and, consequently, shows that these sets of measurements posses a form of hidden nonlocality which can be revealed in such a scenario. We furthermore prove that the maximum amount of nonlocality that can be extracted in such a way is limited by the degree of incompatibility of the given set of measurements, thus effectively establishing an achievable upper bound. In order to obtain these results, we introduce a new object which we term a \emph{generalised set of measurements}, which provides a unifying way to study any scenario involving quantum inputs. 
\end{abstract} 

\maketitle

\begin{textblock*}{3cm}(16.7cm,-11.5cm)
  \footnotesize YITP-24-182
\end{textblock*}
\vspace{-0.3cm}

The study of nonclassical phenomena has remained an intriguing aspect of the theory of quantum mechanics since the formalisation of the theory itself. This includes the study of quantum properties like: entanglement \cite{review_entanglement}, Bell-nonlocality \cite{review_nonlocality, book_nonlocality}, EPR-steering \cite{review_steering1, review_steering2}, measurement incompatibility \cite{review_incompatibility}, contextuality \cite{review_noncontextuality}, discord \cite{review_discord}, coherence \cite{review_coherence}, amongst many others \cite{review_correlations}. Most of these properties are not completely independent from each other. In particular, Bell-nonlocality, entanglement, and incompatibility, are intimately connected.

On the one hand, the presence of Bell-nonlocality, at the level of the correlations, witnesses the presence of entanglement at the level of the shared quantum state \cite{review_nonlocality, review_entanglement}. It is also known that \textit{all} entangled \textit{pure} states can lead to the generation of Bell-nonlocality \cite{Gisin_1991, Gisin_Peres_1992, Popescu_Rohrlich_1992}. Not all entangled states however can lead to Bell-nonlocality, since there exist \textit{mixed} entangled states whose correlations admit a description in terms of local hidden variable (LHV) models and, consequently, cannot be used to violate any Bell-inequality \cite{EL1}. These entangled states, known as \textit{entangled-local} states, have been the subject of exhaustive research over the past three decades \cite{EL1, EL2, review_EL, UN1, UN2}. In particular, great effort has been invested into finding mechanisms under which entangled-local states can still potentially be used to generate or activate nonlocality. Examples of such mechanisms include: local filtering operations \cite{LF1, LF2, LF3, LF4, LF5, LF6, LF7, LF8, LF9, LF10}, activation via tensoring of two different states \cite{NV}, tensoring with the same state (also called superactivation) \cite{SA1, SA2}, distribution of the state in networks \cite{networks}, activation via a scenario with quantum inputs \cite{BNL}, amongst various other approaches \cite{broadcasting, YCL1, YCL2, review_nonlocality, book_nonlocality, review_ANL}.

On the other hand, the presence of Bell-nonlocality also guarantees that the sets of measurements being implemented are incompatible \cite{review_nonlocality, review_incompatibility}. In this regard, it is known that incompatibility of \textit{projective} measurements always leads to Bell-nonlocality in the two-input two-output setting \cite{Wolf_PerezGarcia_Fernandez_2009}. This equivalence stops holding true however when going beyond projective measurements \cite{Priya_2023}, or beyond the two-input two-output scenario, as there exist sets of incompatible measurements which cannot lead to the violation of any Bell-inequality \cite{IL0, IL1, IL2}. These sets of measurements can then be addressed as being \emph{incompatible-local}, in analogy with the counterpart for states. Whilst measurement incompatibility is considered as a valuable resource in its own right \cite{review_incompatibility}, it is particularly desirable for it to also be able to generate nonlocal correlations, as these can  in turn be used to fuel quantum information processing tasks, like semi and fully device independent protocols \cite{book_nonlocality}. Similarly to the case for states, it is then desirable to develop mechanisms under which these incompatible-local sets of measurements can still be used to generate nonlocality. This poses the question of whether incompatible sets of measurements can display forms of hidden nonlocality. In this regard, whilst it has recently been argued that there does not seem to be a reasonable analogue of local filtering operations for sets of measurements \cite{Hsieh_2023, Hsieh_2022}, it nevertheless has been proven that \textit{all} sets of measurements can generate nonlocality in a multi-qubit scenario, where all parties implement the same set of measurements, and therefore displaying a type of superactivation \cite{Plavala_2024}. The scenarios for states do not straightforwardly extend to measurements and, consequently, it becomes of relevance to explore the intricacies of revealing the hidden nonlocality contained in sets of measurements.

In this paper we explore the hidden nonlocality of incompatible sets of measurements in the scenario where quantum inputs are considered. This scenario was first introduced by F. Buscemi in the context of quantum states \cite{BNL}, and so we refer to it here as the Buscemi scenario, and the nonlocality extracted in such a way as Buscemi nonlocality. Explicitly, this scenario considers \emph{quantum} inputs in the form of quantum states, in contrast to the \emph{classical} inputs in the standard Bell scenario. Working within this scenario, we prove that \emph{all} incompatible sets of measurements can be used to generate nonlocality.  In particular, our results hold true for the above-mentioned incompatible-local sets of measurements \cite{IL0, IL1, IL2} and, therefore, show that these apparently useless sets of measurements can still be used to generate nonlocality, when appropriately implemented within the Buscemi scenario. 

The rationale behind our construction follows from implementing three main ideas. First, we extend the concept of a (\emph{standard})  set of measurements to a more general mathematical object which we call a `\emph{generalised}' set of measurements. Second, we identify that the concept of nonclassical teleportation \cite{NCT1, NCT2, NCT3} establishes a bridge between the incompatibility of generalised sets of measurements and the Buscemi-nonlocality that they can generate. Third, we establish quantitative relations between the resources of incompatibility, nonclassical teleportation, and Buscemi-nonlocality. We do this by employing the theoretical machinery of quantum resource theories (QRTs) \cite{review_QRTs1, review_QRTs2}, in particular by considering the resource quantifier of \emph{robustness of resource} \cite{RoE, GRoE, RoNL_RoS_RoI, RoS, RoA, RoC, RoM, FB1, FB2, FB3, RoNL_RoS_RoI, RoT, RoT2, RT_magic, citeme1, QRT_MO}. The appropriate implementation of these three ideas allows for the derivation of a quantitative relationship between the incompatibility of a standard set of measurements and the maximum amount of Buscemi-nonlocality that it can generate. As a consequence of this general statement, it follows that \emph{all} incompatible sets of measurements can generate Buscemi-nonlocality. This demonstrates for the first time an instance of `generic hidden nonlocality', for arbitrary sets of incompatible measurements, regardless of their dimensionality, rather than for a quantum state.

\section{Standard and generalised measurement sets}

The most general (destructive) measurement that can be performed in quantum theory (known as a  positive operator-valued measure (POVM)) is given by $\mathbb{M}=\{M_a\}$, where $M_{a}\geq 0$, $\forall a=1,...,o_A$, and $\sum_a M_{a} = \mathds{1}$, with $\mathds{1}$ the identity operator. A \emph{standard measurement set} is a set $\{\mathbb{M}_{x}\}$, with $\mathbb{M}_x$ a measurement, $\forall x=1,...,i_A$.  Without loss of generality, all measurements are assumed to have the same number of outcomes $o_A$. We denote the elements of each measurement by $M_{a|x}$ and alternatively write a measurement set as $\{ M_{a|x} \}$. 

We first introduce the concept of a \textit{generalised measurement set} $\{M_{a|\omega_x}^A\}$ as a set of operators defined by
\begin{align} \label{eq:GSM}
    M_{a|\omega_x}^{A}
\coloneqq 
\tr_{A'}
[
M^{\rm A'A}_a
    (
    \omega^{A'}_x \otimes \mathds{1}^A
    )
] \qquad \forall a,x
\end{align}
with $\{N_a^{A'A}\}$ a bipartite measurement (on systems labelled $A'$ and $A$), $\{\omega_x^{A'}\}$ a set of states (density operators), and $\tr_{A'}[\cdot]$ the partial trace over the $A'$ subsystem. 

\begin{figure}[t!]
    \centering
    \includegraphics[scale=1.1]{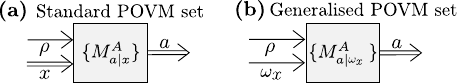}
    \caption{\textbf{(a)} A \textit{standard} POVM set has a classical input $x$ and produces a classical output $a$. \textbf{(b)} A \textit{generalised} POVM set on the other hand has a quantum input (quantum state $\omega_x$) and produces a classical output $a$. The elements of the generalised POVM set $\{M_{a|\omega_x}^A\}$ are considered as 
    $
    M_{a|\omega_x}^{A}
    \coloneqq 
    \tr_V
    [
    M^{VA}_a
        (
        \omega^{V}_x \otimes \mathds{1}^A
        )
    ]
    $, with $\{M_a^{VA}\}$ a bipartite POVM, and $\{\omega_x^V\}$ a set of states. A generalised measurement process takes quantum inputs $\omega_x$, $\rho$, and generates a classical outcome $a$ with probability $p(a|x) = \tr[M_{a|\omega_x} \rho]$.
    }
    \label{fig:fig1}
\end{figure}

A first remark here is that a generalised measurement set can be understood as a ``set of measurements" since it satisfies: $M_{a|\omega_x}^{A} \geq 0$, $\forall \omega_x$, and $\sum_a M_{a|\omega_x}^{A} =\mathds{1}^A$, $\forall \omega_x$. A second remark is that generalised measurement sets are indeed more general than standard measurement sets. In particular, given a (standard) set of measurements $\{L_{a|x}^A\}$, one can always construct a generalised set $\{M_{a|\omega_x}^{A}\}$ which reduces to it, by choosing a controlled measurement $N^{A'A}_a \coloneqq \sum_{x'} \ketbra{x'}{x'}^{A'} \otimes L_{a|x'}^{A}$, and the set of classical states $\omega_x^{A'} \coloneqq \ketbra{x}{x}^{A'}$, with $\{\ket{x}\}$ an orthonormal basis. It is straightforward to check that this generalised measurement set reduces to the standard one, by substituting into \eqref{eq:GSM},  $M_{a|\omega_x}^A = L_{a|x}^A$, $\forall a,x$. Generalised measurements sets can therefore be seen alternatively as resulting from a parent bipartite measurement $M_a^{A'A}$ applied to a controlled state of the form $\omega_x^{A'} \ot \mathds{1}^A$, where the control register $A'$ is a quantum system instead of a classical one. We can also think about this as part of a generalised measurement process that takes quantum inputs $\omega_x$, $\rho$, and generates a classical outcome $a$ with probability $p(a|x) = \tr[M_{a|{\omega}_x} \rho]$ \cite{Schmid1, Schmid2, Schmid3}. In \autoref{fig:fig1} we pictorially contrast standard and generalised measurement sets.

We now generalise the notion of \emph{measurement incompatibility} to these more general sets of measurements. 

\section{Incompatibility of generalised measurement sets}

The notion of incompatibility of a set of quantum measurements dates back to the early days of quantum theory \cite{I1, I2, I3, I4}, and it is still a subject of intensive research \cite{T1, T2, T3, T4, T5, T6, T7}. A (standard) set of measurement $\{M_{a|x}\}$ is said to be \emph{compatible} when it is possible to perform a \emph{single} measurement in place of the individual measurements. More precisely, a set of measurements is compatible when there exists a parent measurement $\mathbb{G}=\{G_\lambda\}$  and a conditional probability distribution $p(a|x,\lambda)$, such that $M_{a|x}=\sum_\lambda p(a|x,\lambda) G_\lambda$, $\forall a,x$ \cite{review_incompatibility}. This says that the whole set of measurements can be reproduced from the single measurement $\mathbb{G}$, via classical (probabilistic) post-processing.

The notion of incompatibility can naturally be extended to generalised measurement sets, in direct analogy to its standard counterpart. We will refer to this as `generalised incompatibility'. As in the above, there is a single parent measurement $\mathbb{G}=\{G_\lambda\}$ that is performed on the incoming quantum system. However, whereas previously the measurement choice (input) $x$ was classical, and was used along with $\lambda$ to generate the outcome $a$ according to $p(a|x,\lambda)$, since the input is now a quantum state $\omega_x^{A'}$, we must use a measurement $\mathbb{H}_\lambda = \{H_{a|\lambda}\}$ (conditional on the outcome $\lambda$ of the parent measurement) to generate the distribution $p(a|x,\lambda) = \tr[H^{A'}_{a|\lambda} \omega_x^{A'}]$.  That is, a generalised set of measurements $\{M_{a|\omega_x}^A\}$ is compatible if it can be written as
\begin{equation}
\label{e:gen parent}
M_{a|\omega_x}^A 
=
\sum_\lambda 
\tr
[
H_{a|\lambda}^{A'}
\omega_x^{A'}
]
G_\lambda^A 
,
\qquad 
\forall a,\omega_x
,
\end{equation}
for suitable measurements $\mathbb{G}$ and $\{\mathbb{H}_\lambda\}$. If it is not possible to write a generalised measurement set in the form \eqref{e:gen parent}, then we will say that it is an \emph{incompatible} generalised measurement set. 

\section{Parallels to Bell-nonlocality and EPR-steering}

With the above definition in place, we will now show that we obtain exact parallels of the relationship that exists between the compatibility of sets of (standard) measurements and their ability to produce Bell-nonlocal correlations or EPR-steering. 

In a (bipartite) Bell scenario, two parties are each given a classical input, $x$ and $y$ respectively, and use this to choose from one of a set of measurements, $\{M_{a|x}^A\}$ and $\{M_{b|y}^B\}$ respectively, which they perform on a shared quantum state $\rho^{AB}$, leading to a set of joint conditional probabilities $p(a,b|x,y) = \tr[(M_{a|x}^{A}\otimes M_{b|y}^{B})\rho^{AB}]$, often referred to as a \emph{behaviour}. If Alice uses a set of compatible measurements, then $p(a,b|x,y)$ will have the structure of a \emph{local hidden variable model}, $p(a,b|x,y) = \sum_\lambda q(\lambda) p(a|x,\lambda) p(b|y,\lambda)$, and hence cannot violate a Bell inequality and does not exhibit nonlocality. This can be seen by direct substitution, with $p(a|x,\lambda)$ the classical post-processing of Alice's compatible set of measurements, $q(\lambda) = \tr[(G_\lambda^A \otimes \mathds{1}^B)\rho^{AB}]$, $\rho_\lambda^B = \tr_A[(G_\lambda^A \otimes \mathds{1}^B)\rho^{AB}]/q(\lambda)$ and $p(b|y,\lambda) = \tr[M_{b|y} \rho_\lambda^B]$. That is, it is \emph{necessary} that both Alice and Bob use \emph{incompatible measurements} in order to produce nonlocal correlations. 

We will now see that we arrive at an identical structure when considering generalised measurement sets and Buscemi nonlocality. In a (bipartite) Buscemi scenario, two parties are each given a \emph{quantum input}, $\omega_x^{A'}$ and $\zeta_y^{B'}$ respectively. As in the Bell scenario, they also share a quantum state $\rho^{AB}$. They each perform a (fixed) measurement (of their choosing) $M_a^{A'A}$ and $M_b^{BB'}$ on their pair of quantum states, leading to a set of joint conditional probabilities (behaviour) $p(a,b|\omega_x,\zeta_y) = \tr[(M_a^{A'A} \otimes M_b^{BB'})(\omega_x^{A'} \otimes \rho^{AB} \otimes \zeta_y^{B'})]$. As a first observation, using the definition \eqref{eq:GSM}, we can rewrite this as $p(a,b|\omega_x,\zeta_y) = \tr[(M_{a|\omega_x}^A\otimes M_{b|\zeta_y^B})\rho^{AB}]$, i.e.~in a form which is a direct analogue of the form from Bell nonlocality, with $M_{a|\omega_x}$ replacing $M_{a|x}$ and $M_{b|\zeta_y} = \tr_{B'}[M_b^{BB'}(\mathds{1}^B \otimes \zeta_y^{B'})]$ replacing $M_{b|y}$ (and we note that we have reordered the systems for Bob compared to Alice, purely for notational convenience). If we now assume that Alice's generalised measurement is compatible, we see that the behaviour $p(a,b|\omega_x,\zeta_y)$ has the form
\begin{equation}
\label{e:Buscemi local}
p(a,b|\omega_x,\zeta_y) = \sum_\lambda q(\lambda) \tr[H_{a|\lambda}^{A'}\omega_x^{A'}]\tr[H_{b|\lambda}^{B'}\zeta_y^{B'}],
\end{equation}
where as above, $q(\lambda) = \tr[(G_\lambda^A \otimes \mathds{1}^B)\rho^{AB}]$ and $\rho_\lambda^B = \tr_A[(G_\lambda^A \otimes \mathds{1}^B)\rho^{AB}]/q(\lambda)$, but now $H_{b|\lambda}^{B'} = \tr_B[M_b^{BB'} ( \rho_\lambda^{B} \otimes \mathds{1}^{B'})]$ can be viewed as a generalised measurement. The form \eqref{e:Buscemi local} is a \emph{local model} in the Buscemi scenario; it says that each party just performs a local (but classically correlated) measurement on their respective quantum input, in order to generate the correlations. Thus, we arrive at the same structure as for Bell-nonlocality: it is necessary that both Alice and Bob use \emph{incompatible generalised measurement sets} in order to produce Buscemi nonlocality. 

We also arrive at the same parallel if we consider the phenomena of EPR-steering and teleportation. In particular, in a (bipartite) EPR-steering scenario, in contrast to the Bell scenario, it is now only Alice who receives an input, and it is the correlations between Alice's input and measurement outcome, and the corresponding state that she remotes prepares for Bob (conditional on her outcome) that is of interest. It is convenient to consider the \emph{unnormalised} conditional states prepared, which are succinctly given by $\sigma_{a|x}^B = \tr_A[(M_{a|x} \otimes \mathds{1}^B)\rho^{AB}]$, with the collection referred to as an \emph{assemblage}. If Alice performs a compatible set of measurements, then the conditional states take the form $\sigma_{a|x} = \sum_\lambda q(\lambda) p(a|x,\lambda) \rho_\lambda^B$, where all terms on the right-hand-side are as previously. This form is known as a \emph{local hidden state} (LHS model) model, which is unable to violate a steering inequality and does not exhibit EPR steering. 

Let us now turn our attention to the scenario of quantum teleportation, focusing on a case where Alice isn't given a completely arbitrary unknown state, but is rather given an unknown state $\omega_x$ from a predetermined set of states. Alice shares an (entangled) state with Bob $\rho^{AB}$, and performs a joint measurement $M_a^{A'A}$ on her two particles. Traditionally in teleportation, Alice then communicates her measurement outcome to Bob, however we can focus exclusively on the first stage of a teleportation protocol, which is the `nonlocal' part. What is of interest are the \emph{conditional} states prepared for Alice, depending on the state which was given to teleport $\omega_x$, and her measurement outcome $a$. As in EPR-steering, it is convenient to consider the \emph{unnormalised} conditional states, which are succinctly given by $\tau^B_{a|\omega_x} = \tr_{A'A}[(M_{a}^{A'A} \otimes \mathds{1}^B)(\omega_x^{A'}\otimes \rho^{AB})]$. This collection of states is sometimes referred to as a \emph{teleportation assemblage}. 

Again, as a first observation, using the definition of a generalised measurement \eqref{eq:GSM}, we can re-express a teleportation assemblage as $\tau_{a|\omega_x}^B = \tr_A[(M_{a|\omega_x}^A \otimes \mathds{1}^B)\rho^{AB}]$, i.e.~in a form identical to the form of an (EPR-steering) assemblage, with $M_{a|x}$ replaced by $M_{a|\omega_x}$. As previously, if we now consider a \emph{compatible} set of generalised measurements, then the teleportation assemblage has the form 
\begin{equation}
\label{e:teleportation model}
\tau_{a|\omega_x}^B = \sum_\lambda q(\lambda) \tr[H_{a|\lambda}^{A'}\omega_x^{A'}] \rho_\lambda^B,
\end{equation} 
where again all terms on the right-hand-side are as before. This form is again the `local model' in the context of non-classical teleportation \cite{NCT1, NCT2, NCT3}; it can be interpreted as Bob being given the state $\rho_\lambda$, according to $q(\lambda)$, with Alice performing a (classically correlated) measurement on the particle she is given to `teleport' to Bob. Thus, it is necessary that Alice uses a set of incompatible generalised measurements in order to demonstrate non-classical teleportation, in direct analogy to EPR-steering. 

\section{Robustness of incompatibility of generalised measurement sets and its relation to other robustness measures}

We now begin our quantitative analysis of the incompatibility of generalised sets of measurements, which is essentially employing the framework of quantum resource theories \cite{review_QRTs1, review_QRTs2}. That is, we consider generalised measurement sets as the \emph{objects} of the resource theory, with the property to be considered a \emph{resource} being their incompatibility, since in the above we have seen this is necessary in order to lead to Buscemi nonlocality or non-classical teleportation. We will focus exclusively on a well-studied resource quantifier, the \emph{generalised robustness}. This quantifies the minimal amount of `worst case noise' that needs to be mixed with the object of interest, before it ceases to be resourceful. In the present context of a generalised measurement set $\{M_{a|\omega_x}\}$, this noise would be another set of generalised measurements $\{N_{a|\omega_x}\}$  (with the \emph{same} set of quantum inputs $\omega_x$), be it compatible or incompatible, that when mixed with $M_{a|\omega_x}$, would make them compatible, namely having the structure of \eqref{e:gen parent}. Formally, 
\begin{align}
\label{eq:RoI0}
    {R_\mathrm{I}}
    (
    \{ 
    M_{a|\omega_x}^{A}
    \}
    )  
    &=
    \min 
    \quad r, \\
    \text{s.t.}&
    \hspace{0.4cm}
    \frac{M_{a|\omega_x}^{A} 
    +
    r
    N_{a|\omega_x}^{A}}{1+r} 
    = 
    \sum_\lambda \tr[H_{a|\lambda}^{A'}
    \omega_x^{A'}] G_\lambda^A, 
    \nonumber
\end{align}
where the minimisation is over $r \geq 0$, the generalised measurement set $\{N_{a|\omega_x}\}$, the measurement $\mathbb{G}$ and the set of measurements $\{\mathbb{H}_\lambda\}$. 

We note that this reduces to the usual robustness of incompatibility of \emph{standard} sets of measurements  when $\omega_x = \ket{x}\bra{x}$ is a set of orthogonal states.  This connection between standard measurement sets and their generalised counterpart will be useful when proving our main result. 

In a completely analogous fashion it is possible to define generalised robustness measures for non-classical teleportation and Buscemi nonlocality. In particular, the \emph{(generalised) robustness of teleportation} of a teleportation assemblage $R_\mathrm{T}(\{\tau_{a|\omega_x}\})$ is defined analogously to \eqref{eq:RoI0}, except that the `noise' is replaced by an arbitrary teleportation assemblage, and the right-hand-side is replaced by the `local model' \eqref{e:teleportation model}. Similarly, the \emph{(generalised) robustness of Buscemi nonlocality} of a behaviour $R_\mathrm{BN} \left( \{p(a,b|\omega_x,\zeta_y)\} \right)$ is also defined analogously, with the `noise' an arbitrary behaviour (with the \emph{same} quantum inputs $\omega_x$ and $\zeta_y$), and the right-hand-side replaced by the local behaviour \eqref{e:Buscemi local} (details on robustness measures in \cref{a:a0}). 

We are now ready to state our central ideas, which are to use nonclassical teleportation as a ``bridge" between the incompatibility of generalised measurements sets and Buscemi nonlocality. Our first main result is to quantitatively connect the incompatibility of generalised measurement sets to nonclassical teleportation. 
 
\begin{result} \label{r:r1}
    The maximal attainable robustness of teleportation among all shared quantum states $\rho^{AB}$, using a fixed generalised measurement set $\{M_{a|\omega_x}^{A}\}$ is
    \begin{align}\label{eq:r1}
        \max_{\rho^{AB}} R_\mathrm{T}(\{\tau_{a|\omega_x}^{B} \}) = R_\mathrm{I}(\{ M_{a|\omega_x}^{A} \}).\
    \end{align}
\end{result}
\noindent The proof of this result is in \cref{a:a1}. An immediate consequence is that \emph{all incompatible generalised measurement sets can demonstrate nonclassical teleportation}. Our second main result is to further connect nonclassical teleportation to Buscemi nonlocality.
\begin{result}\label{r:r2}
    The maximum attainable robustness of Buscemi nonlocality among all generalised measurement sets of Bob $\{M_{b|\zeta_y}\}$, using a fixed generalised measurement set for Alice $\{M_{a|\omega_x}^{A}\}$ and a fixed shared quantum state $\rho^{AB}$ is
	\begin{align}\label{eq:r2}
		\max_{\{M_{b|\zeta_y}^B\}} R_\mathrm{BN}
        \left(
        \{
        p(a,b|\omega_x,\zeta_y)
        \}
        \right) = 
		R_\mathrm{T}(\{ \tau_{a|\omega_x}^B \}),
	\end{align}
	where $\tau_{a|\omega_x}^B = \tr_A[(M_{a|\omega_x}^A \otimes \mathds{1}^B)\rho^{AB})$ is the teleportation assemblage generated by $\{M_{a|\omega_x}^{A}\}$ and $\rho^{AB}$.
\end{result}
\noindent The proof of this result is in \cref{a:a2}. An immediate consequence is that \emph{all teleportation assemblages can demonstrate Buscemi-nonlocality}. These two results together imply the following: 

\begin{result}\label{c:c1}
    The maximum attainable robustness of Buscemi nonlocality among all generalised measurement sets of Bob $\{M_{b|\zeta_y}^B\}$ and all shared quantum states $\rho^{AB}$, using a fixed generalised measurement set $\{M_{a|\omega_x}^{A}\}$ is
    {\small\begin{align}
    \max_{\rho^{AB}}
    \max_{\{M_{b|\zeta_y}^B\}}
    R_\mathrm{BN}
    \left(
    \{
    p(a,b|\omega_x,\zeta_y)
    \}
    \right) 
    = 
    R_\mathrm{I}
    (
    \{
    M_{a|\omega_x}^{A} 
    \}
    ),
    \end{align}}
\end{result}
Finally, we also have the following result:
\begin{result} All sets of incompatible measurements can generate Buscemi nonlocality.
\end{result}
\begin{proof}
	For any incompatible (standard) set of measurements $\{M_{a|x}^A\}$ we have
	\begin{align}
		0
		&\overset{1}{<}
		R_\mathrm{I}
		(
		\{M_{a|x}^A \}
		),\nonumber\\
		&\overset{2}{=}
		R_\mathrm{I}(
		\{M_{a|\ket{x}\bra{x}}^{A} \}
		),\nonumber\\
		&\overset{3}{=}
		\max_{
			\rho^{AB}
		}
		\max_{
			\{
			M_{b|\zeta_y}^B 
			\}
		}
		R_\mathrm{BN}
		\left( 
		\left\{p(a,b|\ket{x}\bra{x},\zeta_y)\right\}
		\right).
	\end{align}
    The first line follows since, by assumption the set of measurements $\{M_{a|x}^A\}$ is incompatible, and the (generalised) robustness of incompatibility is a faithful measure \footnote{A faithful resource measure achieves the value zero if and only if the object belongs to the free set}. The second line follows by viewing a standard set of measurements as a generalised set,  with orthogonal quantum inputs. The third line follows from \cref{c:c1}. This shows that the robustness of Buscemi-nonlocality is strictly greater than zero, and consequently, the incompatible set $\{M_{a|x}^A \}$ must therefore be able to generate nonlocality in the Buscemi scenario, with the maximum amount of nonlocality that can be extracted being upper bounded by the amount of incompatibility measured by the generalised robustness. 
\end{proof}

We emphasise that these results apply to \emph{any} incompatible (standard) set of measurements, including measurement sets that are incompatible-local. This result therefore establishes a general scenario for the activation of nonlocality of incompatible measurements. A diagrammatic representation of our results in terms of the robustness quantifier as well as a comparison between our findings and previous works in the literature can be seen in Fig. \ref{fig:fig2} (\cref{a:a0}). In particular, it is relevant to compare the results of this work to those in \cite{Plavala_2024}, so we do this in what follows.

The main result in \cite{Plavala_2024} establishes that \textit{all} incompatible sets of measurements \emph{for qubits} can generate Bell-nonlocality, in a scenario where the parties share a multi-qubit quantum state, and all the parties implement the same set of measurements.  The setting explored in this paper on the other hand, involves a nonlocality scenario involving quantum inputs. Specifically, the nonlocal correlations are here generated via a shared bipartite state (not necessarily restricted to qubit systems), whilst the measurements in question are considered in a setup with quantum inputs (as opposed to many copies of the measurements). The main result here also holds for \textit{all} incompatible sets. Having highlighted these differences between the approach explored here and that in \cite{Plavala_2024}, we can now also remark on some similarities. 

We remark that both approaches share the same motivation of generating nonlocality using \emph{all} incompatible measurements, incompatible-local POVMs in particular, and thus proposing ways of revealing the hidden nonlocality contained within sets of measurements. From a technical point of view, a specific aspect in common is that both approaches employ somewhat similar mathematical techniques. The case for superactivation employs the hyperplane separation theorem, whilst the scenario considered here employs semidefinite programming (SDP) techniques, or conic programming more generally. The resource measures considered in this work can be thought of as a refinement of the hyperplane separation theorem, and so it can be argued that both approaches have this in common. It would be interesting to explore whether this is just a matter of coincide, or if both approaches can somehow be connected at a more fundamental level.

\section{Conclusions}

In this paper we have proven that \emph{all} incompatible sets of measurements can be used to generate nonlocality in a scenario involving quantum inputs. Explicitly, in a quantitative manner, we proved that the maximum amount of nonlocality that can be extracted in such a way, measured in terms of the generalised robustness of resource, is upper bounded by the amount of incompatibility in the set of measurements, and furthermore that this is an achievable upper bound. These results apply, in particular, to \emph{all} incompatible-local sets of measurements and therefore, effectively revealing a form of nonlocality hidden within incompatible-local POVM sets. This shows that, although apparently useless at first sight, incompatible-local sets of measurements still happen to posses a form of hidden nonlocality, which can be extracted and exploited for the benefit of information-processing protocols. These results are experimentally-friendly and can potentially be tested with current technology, like bulk-optics or photonic setups. Finally, a key ingredient in the derivation of these results was the introduction of the mathematical object of (sets of) generalised measurements, and so it would be interesting to explore further scenarios where these objects play a central role, as well as their potential application to information-theoretic operational tasks \cite{OT1, OT2, OT3, OT4, OT5, OT6, OT7}.


\emph{Acknowledgements.---}We thank Tom Purves, Ben Jones, and Roope Uola for insightful discussions. A.F.D. thanks Ryo Takakura, Ivan \v{S}upi\'{c}, and Erkka Haapasalo for insightful discussions on measurement incompatibility. A.F.D. acknowledges support from the International Research Unit of Quantum Information, Kyoto University, the Center for Gravitational Physics and Quantum Information (CGPQI), and COLCIENCIAS 756-2016. P.L.-B. acknowledges funding from Polish National Agency for Academic Exchange (NAWA) through grant BPN/PPO/2023/1/00018/U/00001. C.E.S. acknowledges support from University of C\'ordoba (Grants FCB-12-23). Part of this work was carried out whilst A.F.D. was a PhD student in the Quantum Engineering Centre for Doctoral Training (QE-CDT). Part of this work was carried out while P.S. was a Royal Society URF (NFQI) and a CIFAR Azrieli Global Scholar.

\bibliographystyle{apsrev4-1}
\bibliography{bibliography.bib}

\appendix

\section{Resource measures for incompatibility, teleportation, and Buscemi nonlocality}
\label{a:a0}

Let $\{M_{a|\omega_x}^{A}\}$ be a generalised POVM set, the generalised robustness of incompatibility is given by:
\begin{align}
    {R_\mathrm{I}}
    (\{ M_{a|\omega_x}^{A}\}
    )  
    &\coloneqq 
    \min_{
        r\geq 0, 
        \,
        p_{A|X\Lambda},
        \,
        \mathbb{G}^A
        ,
        \{N_{a|\omega_x}^{A}\}
    }
    \quad r,
    \label{eq:RoI}
\end{align}
with the minimisation over all conditional probability mass functions (PMFs) $p_{A|X\Lambda}$, all POVMs $\mathbb{G}^A = \{G_{\lambda}^{A}\}$, all generalised POVM sets $\{N_{a|\omega_x}^{A}\}$ of the form \eqref{eq:GSM}
which satisfy:
{\small \begin{align}
    M_{a|\omega_x}^{A} 
    +
    r 
    N_{a|\omega_x}^{A} 
    &= 
    (1+r) 
    \sum_{\lambda} 
    p(a|x,\lambda)  
    G^A_{\lambda},
    \hspace{0.2cm}
    \forall\,a,\omega_x.
    \label{eq:RoI1}
\end{align}}
We now consider the generalised robustness of nonclassical teleportation of a teleportation assemblage $\{\tau_{a|\omega_x}^{B}\}$ as a quantifier for nonclassical teleportation:
\begin{align}
    {R_\mathrm{T}}
    (
    \{\tau_{a|\omega_x}^{B} \}
    ) 
    \coloneqq
    \min_{
        r\geq 0,
        \,
        p_{A|X\Lambda},
        \,
        p_\Lambda,
        \,
        \{\sigma_\lambda^B\}
        ,
        \{\sigma_{a|\omega_x}^{B}\}
    }
    \quad r,
    \label{eq:RoT}
\end{align}
with the minimisation over all conditional PMFs $p_{A|X\Lambda}$, all state assemblages $p_{\Lambda}, \{\sigma_\lambda^B\}$ of $\rho^{AB}$, all teleportation assemblages $\{\sigma_{a|\omega_x}^{B}\}$, $\sigma_{a|\omega_x}^{B}= \tr_{A}[(N_{a|\omega_x}^{A} \ot \mathds{1}^{B})\sigma^{AB}]$, with  arbitrary generalised POVM sets $\{N_{a|\omega_x}^{A}\}$, arbitrary bipartite states $\sigma^{AB}$, $\sigma^B = \tr_A (\rho^{AB})$, and such that these variables satisfy:
{\small \begin{align}
    \tau_{a|\omega_x}^B 
    +
    r
    \sigma_{a|\omega_x}^B
    &=
    (1+r)  \sum_{\lambda}
    p(a|x,\lambda) 
    \,
    p(\lambda)
    \,
    \sigma_{\lambda}^B,
    \hspace{0.3cm} 
    \forall\,a, x
    .
\end{align}}
The generalised robustness of nonlocality of a Buscemi behaviour $\left \{p(a,b|\omega_x,\omega_y) \right\}$ is given by:
\begin{align}
    \hspace{-1.5cm}
    {R_\mathrm{BN}}
    \left( 
    \left\{p(a,b|\omega_x,\omega_y)\right\}
    \right)
    \coloneqq
    \min_{
        r \geq 0, 
        \,
        p_{A|X\Lambda},
        \,
        p_{B|Y\Lambda},
        \,
        p_\Lambda,
        \,
        p^G_{AB|XY}
    }
    \hspace{-1.5cm}
    r
    \label{eq:RRoBN}
\end{align}
with the minimisation over all conditional PMFs $p_{A|X\Lambda}$, $p_{B|Y\Lambda}$, $p^G_{AB|XY}$, all PMFs $p_\Lambda$, and such that these variables satisfy the following condition $\forall a,b,x,y$:
\begin{multline}
    p(a,b|\omega_x,\omega_y)
    +
    r\,
    p^G(a,b|x,y)
    \\
    =
    (1+r)
    \sum_\lambda 
    p_A(a|x,\lambda)
    \,
    p_B(b|y,\lambda)
    \,
    p(\lambda)
    .
    \label{eq:RoBNc1}
\end{multline}
The PMF $p_A$ can be understood as $p_A(a|x,\lambda) := \tr[H_{a|\lambda}^{A'}\omega_x^{A'}]$, and similarly for $p_B$, distinguishing between auxiliary systems $A'$ and $B'$, respectively.

\begin{figure}[h!]
    \centering
    \includegraphics[scale=1]{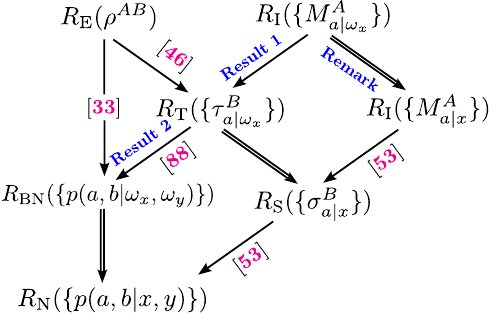}
    \caption{Hierarchy of various quantum resources via geometric quantifiers. The quantifiers depicted are the robustness of: entanglement ($R_\mathrm{E}$), incompatibility ($R_\mathrm{I}$), teleportation ($R_\mathrm{T}$), EPR-steering ($R_\mathrm{S}$), Bell-nonlocality ($R_\mathrm{N}$), and Buscemi-nonlocality ($R_\mathrm{BN}$). The single line arrow notation $R_\mathrm{E}(\rho^{AB}) \rightarrow R_\mathrm{T} (\{\tau^{B}_{a|\omega_x}\})$ means that $R_\mathrm{E}(\rho^{AB}) \geq R_\mathrm{T}(\{\tau^{B}_{a|\omega_x}\})$, $\forall \{\omega_x^{V}\}$, $\forall \{M_a^{VA}\}$, and similarly for the other quantities. The double line arrow notation 
    $
    R_\mathrm{I}
    (
    \{M^{A}_{a|\omega_x}\}
    ) 
    \Rightarrow
    R_\mathrm{I}
    (
    \{M^{A}_{a|x}\}
    )
    $ means the left quantity reduces to the one in the right when considering classical inputs, and similarly for the other quantities. The connection between entanglement and  nonlocality with quantum inputs (Buscemi nonlocality) was first introduced in \cite{BNL} by F. Buscemi. This connection can alternatively be divided by first connecting entanglement to nonclassical teleportation \cite{NCT1}, followed by connecting nonclassical teleportation to Buscemi nonlocality (\cref{r:r2}, see also \cite{MDI16}). The quantitative relations between Bell-nonlocality, EPR-steering, and incompatibility were first established in \cite{RoNL_RoS_RoI}. In this work we introduce the concept of generalised POVM sets, and establish the connection between their incompatibility and nonclassical teleportation (\cref{r:r1}). This connection then effectively establishes a bridge between incompatibility and Buscemi nonlocality, from which it follows that all incompatible POVM sets can generate nonlocality in the Buscemi scenario.
    }
    \label{fig:fig2}
\end{figure}

\section{Proof of \cref{r:r1}}
\label{a:a1}

We want to prove that the maximal attainable nonclassical teleportation of arbitrary teleportation assemblages $\{\tau_{a|\omega_x}^{B}\}$ employing a fixed generalised POVM set $\{M_{a|\omega_x}^{A}\}$ is given by:
\begin{align}
    \max_{
    \rho^{AB}
    }
    R_\mathrm{T}
    (
        \{\tau_{a|\omega_x}^{B} \}
    )
    = 
    R_\mathrm{I}
    (
        \{ M_{a|\omega_x}^{A} \}
    ),
    \label{eq:r1a}
\end{align}
with the maximisation over all quantum states. Let us begin by showing that:
\begin{align}
    \label{rot_geq_roi}
    \max_{
    \rho^{ AB}
    } 
    R_\mathrm{T}
    (
    \{\tau_{a|{\omega_x}}^B\}
    ) 
    \geq  
    R_\mathrm{I} 
    (
    \{M_{a|\omega_x}^A\}
    ).
\end{align}

\begin{proof}(of \eqref{rot_geq_roi})
Fix a generalised POVM set $\{M_{a|\omega_x}^{A}\}$, and consider a potentially suboptimal guess for the optimisation in the LHS model of \eqref{rot_geq_roi} as $\rho^{AB} = \phi_+^{AB} \coloneqq  d^{-1} \sum_{i,j} \ketbra{i}{j}^A \ot \ketbra{i}{j}^B$. The teleportation assemblage can then be written as:
\begin{align}
    \tau_{a|{\omega_x}}^{ B} 
    =
    \tr_{\rm A}
    [
    (
    M_{a|\omega_x}^{ A} 
    \ot
    \mathds{1}^{ B}
    )
    \phi_+^{ AB}
    ] 
    =
    \frac{1}{d}
    (
    M_{a|\omega_x}^{ B}
   )^T. 
\end{align}
The last equality follows because, for any $O^A$ we have:
\begin{align}
    &\tr_{A}
    \left[
    (O^{A} \ot \mathds{1}^{B})
    \phi_+^{AB}
    \right]
    \nonumber
    \\
    &= \frac{1}{d} 
    \sum_{i,j} 
    \tr_{\rm A}
    \left[
    (
    O^{\rm A} 
    \ot
    \mathds{1}^{\rm B}
    )\ketbra{i}{j}^{\rm A} 
    \ot
    \ketbra{i}{j}^{\rm B}
    \right]
    \nonumber
    \\
    &= \frac{1}{d} \sum_{i,j} \braket{j|O^{\rm A}|i} \ketbra{i}{j}^{\rm B} 
    \nonumber
    \\
    &= \frac{1}{d} \sum_{i,j} O_{ji}^{\rm A} \ketbra{i}{j}^{\rm B}
    \nonumber
    \\
    &= \frac{1}{d} (O^{\rm B})^T
    .
    \label{eq:transp}
\end{align}
The LHS model of (\ref{rot_geq_roi}) can then be lower bounded as: 
\begin{align}
     \max_{\rho^{AB}
     } 
     R_\mathrm{T}
     (
     \{
     \tau_{a|{\omega_x}}^{B}
     \}
     ) 
     \geq & 
     \min_{
        r\geq 0,
        \,
        p_{A|X\Lambda},
        \,
        p_\Lambda,
        \,
        \{\sigma_\lambda^B\}
        ,
        \{\sigma_{a|\omega_x}^{B}\}
        } 
     \hspace{-0.1cm}
     r,
\end{align}
with $r \geq 0$ being subject to:
{\small \begin{align}
    \tau_{a|{\omega_x}}^{B}
    + 
    r\,
    \sigma_{a|{\omega_x}}^{B}
    &= (1+r)\! 
    \sum_{\lambda} 
    p(\lambda)\,
    p(a|x, \lambda)\,
    \sigma_{\lambda}^{ B}, 
    \\
    \sum_{\lambda} 
    p(\lambda) \,
    \sigma_{\lambda}^{B} 
    &=
    \rho^{B}
    =
    \frac{\mathds{1}^B}{d}
    .
\end{align}}
In the last equality we use 
$
\rho^{B} =
\tr_A(\phi^{AB}_+)
= \frac{\mathds{1}^{B}}{d}
$. Taking the transpose of the first line and multiplying it by $d$ we obtain:
{\small\begin{align}
    \hspace{-0.2cm}
    d
    (\tau_{a|{\omega_x}}^{B})^T
    \hspace{-0.1cm}
    +
    r  
    d(\sigma_{a|{\omega_x}}^{B})^T
    \hspace{-0.1cm}
    &= 
    (1+r) 
    \sum_{\lambda} 
    p(a|x,\lambda) 
    \hspace{-0.1cm}
    \left[
    d \, 
    p(\lambda)
    \sigma_{\lambda}^B
    \right]^T
    \hspace{-0.3cm}
    , 
    \\
    \sum_{\lambda}
    p(\lambda)\,
    \sigma_{\lambda}^{B} 
    &= \frac{1}{d} 
    \mathds{1}^{B}.
\end{align}}
Defining $G_{\lambda}^{B} := d \, p(\lambda) (\sigma_{\lambda}^B)^{T}$ and using that the operators
$
d(\{\sigma_{a|\omega_x}^B\})^T
$ define a valid generalised POVM set, we can rewrite the above set of constraints as:
\begin{align}
    M_{a|\omega_x}^{B} 
    + 
    r 
    N_{a|\omega_x}^{B} 
    &= (1+r) \sum_{\lambda} 
    p(a|x, \lambda) 
    G_{\lambda}^{B},
    \label{r1_pr_11}
    \\ 
    \sum_{\lambda} 
    G_{\lambda}^{B} 
    &= \mathds{1}^{B}, \quad G_{\lambda}^{B} \geq 0, 
    \quad \forall \, \lambda.
    \label{r1_pr_12}
\end{align}
We now note that these constraints are exactly the same as the ones appearing in the definition of $R_\mathrm{I}(\{M_{a|\omega_x}^A\})$ \eqref{eq:RoI1}. The minimal value of $r$, given constraints \eqref{r1_pr_11} and \eqref{r1_pr_12}, is then equal to $R_\mathrm{I} (\{M_{a|\omega_x}^A\} )$, proving \eqref{rot_geq_roi}. 
\end{proof}

We now show the reverse inequality, i.e:
\begin{align}
    \label{rot_leq_roi}
    \max_{
    \rho^{AB}
    } 
    R_\mathrm{T}
    (
    \{
    \tau_{a|{\omega_x}}^{B}
    \}
    ) 
    \leq 
    R_\mathrm{I} 
    (
    \{M_{a|\omega_x}^{A}\}
    ).
\end{align}

\begin{proof}(of \eqref{rot_leq_roi}) Fix a generalised POVM set $\{M_{a|\omega_x}^{A}\}$, and let us assume we have solved the problem defining $ R_\mathrm{I} (\{M_{a|\omega_x}^{ A}\})= r^*$ (the minimum in \autoref{eq:RoI}) using variables $\{G_{\lambda}^{^*}\}$, $p^*(a|x, \lambda)$, and $\{N_{a|\omega_x}^*\}$. For any quantum state $\rho^{AB}$, we can now construct a potentially sub-optimal guess for the optimisation problem defining $R_\mathrm{T}(\{\tau_{a|{\omega_x}}^{B}\})$, as:
\begin{align}
    \label{rot_leq_robi_guess}
    &
    p(\lambda)\, 
    \sigma_{\lambda}^{B} 
    =
    \tr_{A}
    [
    (G_{\lambda}^{*} \ot \mathds{1}^{B}) 
    \rho^{AB}
    ],
    \hspace{0.3cm}
    \forall \lambda,
    \\ 
    \nonumber
    &p(a|x,\lambda) 
    =
    p^*(a|x,\lambda)
    ,\hspace{0.3cm}
    \forall a,x,\lambda,
    \\
    &\sigma_{a|\omega_x}^B
    = 
    \tr_{A}
    [
    (
    N_{a|\omega_x}^* 
    \ot 
    \mathds{1}^{B}
    )
    \rho^{AB}
    ] 
    , \forall a,x
    .
    \nonumber
\end{align}
According to the definition \autoref{eq:RoT} for $R_\mathrm{T}(\{\tau_{a|{\omega_x}}^{B}\})$, the above construction allows us to write:
\begin{align}
    R_\mathrm{T}
    (
    \{\tau_{a|{\omega_x}}^{B}\}
    ) 
    \leq 
    r^* 
    =
    R_\mathrm{I} 
    (
    \{M_{a|\omega_x}^{A}\}
    ).
\end{align}
This holds for any $\rho^{AB}$ and therefore, in particular, for the $\rho^{AB}$ achieving the maximisation in the LHS model of (\ref{rot_leq_roi}). It remains to show that (\ref{rot_leq_robi_guess}) is a feasible choice. Notice that, by the definition of $R_\mathrm{I} (\{M_{a|\omega_x}^{A}\})$, we have that $r \geq 0$, $\sum_{a} p(a|x, \lambda) = 1$ for all $x$, $\lambda$ and $\sum_{\lambda} p(\lambda) \sigma_{\lambda}^{B} = \rho^{ B}$. The last condition to check is whether:
\begin{align}
    \label{eq:rot_leq_robi_eq2}
    \tau_{a|\omega_x}^B
    + 
    r\,
    \sigma_{a|\omega_x}^B
    \stackrel{?}{=}   
    (1+r) 
    \sum_{\lambda}
    p(\lambda)\,
    p(a|x, \lambda)\,
    \sigma_{\lambda}^{B}
    .
\end{align}
Since we know that the starred variables $r^*$, $\{G_{\lambda}^{^*}\}$, $p^*(a|x, \lambda)$, and $\{N_{a|\omega_x}^*\}$, are feasible for $R_\mathrm{I} (\{M_{a|\omega_x}^{A}\})$, we can write:
\begin{align}
    M_{a|\omega_x}^{A} 
    + 
    r 
    N_{a|\omega_x}^*
    =
    (1+r) 
    \sum_{\lambda} 
    p^*(a|x, \lambda) 
    G_{\lambda}^{*}
    .
    \label{eq:previous0}
\end{align}
We can now start from the LHS model of (\ref{eq:rot_leq_robi_eq2}) and get: 
{\small \begin{align}
    &
    \tau_{a|\omega_x}^B
    + 
    r 
    \sigma_{a|\omega_x}^B
    \nonumber
    \\ 
    &\overset{1}{=}
    \tr_{A}
    \left[
    \left(
    \left(
    M_{a|\omega_x}^{A} 
    + 
    r 
    N_{a|\omega_x}^*
    \right) 
    \ot 
    \mathds{1}^{B}
    \right)
    \rho^{AB}
    \right] 
    \\ 
    &\overset{2}{=} 
    \tr_{A}
    \left[
    \left(
    \left(
        (1+r) 
        \sum_{\lambda} 
        p^*(a|x, \lambda) 
        G_{\lambda}^{*}
    \right) 
    \ot 
    \mathds{1}^{B}
    \right)
    \rho^{AB}
    \right] 
    \\
    &\overset{3}{=}
    ( 1+r) 
    \sum_{\lambda} 
    p^*(a|x,\lambda) 
    \tr_{A}
    \left[
    (G_{\lambda}^* 
    \ot 
    \mathds{1}^{B})
    \rho^{AB}
    \right] 
    \\ 
    &\overset{4}{=}
    (1+r) 
    \sum_{\lambda}
    p(\lambda)\,
    p(a|x, \lambda)\,
    \sigma_{\lambda}^{B}.
\end{align}}
In the first line we reorganise. In the second line we use \eqref{eq:previous0}. In the third line we reorganise. In the fourth line we use our guess \eqref{rot_leq_robi_guess}. This shows that \eqref{rot_leq_robi_guess} is a feasible choice of optimisation variables, hence proving \eqref{rot_leq_roi}. Combining \eqref{rot_leq_roi} with  \eqref{rot_geq_roi} proves the claim \eqref{eq:r1a}.
\end{proof}

\section{Intermediate steps for \cref{r:r2}}
\label{a:a2int}

\begin{lemma}[RoT - primal] \label{rot_choi}
    Fix a POVM set $\{M_a^{VA}\}$, a quantum state $\rho^{AB}$ and let $\{\omega_x^{\rm V}\}$ be a tomographically-complete set of states. Define Choi-Jamiołkowski (CJ) operators $J_{a}^{\rm V'B}$ via:
    \begin{align}
        \!\! 
        J_a^{\rm V'B}
        & \! 
        \coloneqq
        \!\tr_{\rm VA}
        \left[
            \left(
                \mathds{1}^{\rm V'}
                \!\!\!\ot\! 
                M_a^{\rm VA} 
                \!\ot\! 
                \mathds{1}^{\rm B}
            \right)
            \left(
                \phi_+^{\rm V'V} 
                \!\!\ot\! 
                \rho^{\rm AB}
            \right)
        \right].
        \label{eq:CJ}
    \end{align}
    Then:
    \begin{align}
        R_\mathrm{T}
        \left(
        \{M_a^{VA}\}, 
        \rho^{AB}
        \right) 
        = 
        \min_{\{F_a^{\rm V'B}\}}
        \,\, r,
    \end{align}
    subject to:
    \begin{align*}
        &J_a^{\rm V'B} 
        +
        r
        K_a^{\rm V'B} 
        = 
        (1+r)   
        F_a^{\rm V'B},
        \quad \forall\,a, 
        \\
        &F_a^{\rm V'B}
        \coloneqq
        \sum_{\lambda} 
        p({\lambda})\, 
        N_{a|\lambda}^{\rm V'} 
        \ot
        \sigma_{\lambda}^{\rm B},
    \end{align*}
    with $\big\{N_{a|\lambda}^{\rm V'}\big\}$ a set measurements, $\{\sigma_{\lambda}^{\rm B}\}$ a set of states, $p_\Lambda$ a PMF, and $\{K_a^{\rm V'B}\}$ the CJ operators associated to a general pair ($\sigma^{AB}$, $N_a^{AB}$).
\end{lemma}
We will denote the convex cone of free teleportation operators by $\ftel$. We will also say that a set of teleportation operators $\big\{F_a^{\rm V'B}\big\} \in \ftel$ if and only if there is a PMF $p_\Lambda$, a set of measurements $\big\{N_{a|\lambda}^{\rm V'}\big\}$ and a set of states $\{\sigma_{\lambda}^{\rm B}\}$ such that we have $F_a^{\rm V'B} = \sum_{\lambda} p(\lambda) N_{a|\lambda}^{\rm V'} \ot \sigma_{\lambda}^{\rm B}$, $\forall a$. This definition is equivalent to the one given in \eqref{eq:RoT} under the assumption that input states $\{\omega_x^{\rm V}\}$ form a tomographically-complete set. In other words, $\{F_a^{\rm V'B}\} \in \ftel$ if and only if any teleportage $\{\tau_{{\rm a}|\omega_x}^{\rm B}\}$ arising from the teleportation instrument specified by $\{F_{a}^{\rm V'B}\}$ is free according to \eqref{eq:RoT}. With this in mind we also define the dual cone of $\ftel$ and label it with $\ftel^*$ so that any element $\{W_a^{\rm V'B}\} \in \ftel^*$ satisfies:
\begin{align}
    \sum_{a} 
    \tr
    \left[
    W_a^{\rm V'B}
    F_a^{\rm V'B}
    \right]
    \geq
    0,
    \qquad
    \forall \, 
    \{F_a^{\rm V'B}\} 
    \in
    \ftel. 
\end{align}
With these we can now write the dual form of the optimisation problem from \cref{rot_choi}, i.e:
\begin{lemma}[RoT - dual] 
    \label{rot_choi_dual}
    Fix a POVM set $\{M_a^{VA}\}$, a quantum state $\rho^{AB}$ and let $\{\omega_x^V\}$ be a tomographically-complete set of states. Then, the dual of $R_\mathrm{T}$ can be written as:
    {\small\begin{align}
        R_\mathrm{T}
        (
        \{M_a^{\rm VA}\},
        \rho^{\rm AB}
        ) 
        = 
        \max_{W_{a}^{\rm V'B}} \,\,
        -\sum_{a}
        \tr
        \left[
        J_a^{\rm V'B}
        W_{a}^{\rm V'B}
        \right]
        ,
    \end{align}}
    such that:
    \begin{align}
        & \{W_a^{\rm V'B}\} 
        \in
        \ftel^* \\
        & \sum_{a} 
        \tr
        \left[
        W_{a}^{\rm V'B}
        \right]
        \leq
        o_{\rm A} 
        \cdot
        d
        ,
    \end{align}
\end{lemma}
where $o_{\rm A}$ stands for the number of outcomes $a$. In a similar way we can rewrite our standard definition of $\rm RoBN$ as a function ${\rm RoN}(\{M_a^{VA}\},\{M_b^{BV'}\}, \rho^{AB})$ for the case of tomographically-complete inputs. We obtain:
\begin{lemma}[RoBN - primal]
    \label{robn-primal}
    Let $\{\omega_x^{\rm V}\}$ and $\{\omega_y^{\rm V'}\}$ be tomographically-complete sets of states and let:
    {\small\begin{align}
        M_{ab}^{\rm VV'} 
        \coloneqq 
        \tr_{\rm AB}
        [
            (
                M_a^{\rm VA}
                \ot
                M_b^{\rm BV'}
            )
            (
                \mathds{1}^{\rm V}
                \ot
                \rho^{\rm AB}
                \ot
                \mathds{1}^{\rm V'}
            )
        ].
        \label{eq:buscemimeas}
    \end{align}}
    Then:
    \begin{align}
        &R_\mathrm{BN}
        (
        \{M_a^{VA}\},
        \{M_b^{BV'}\},
        \rho^{AB}
        ) = 
        \min_{\{F_{ab}^{\rm VV'}\}}
        \, r,
        \label{eq:RRoBNa}
    \end{align}
    subject to:
    \small{\begin{align}
    & 
    M_{ab}^{\rm VV'} 
    + 
    r 
    L_{ab}^{\rm VV'}  
    = 
    (1+r) 
    F_{ab}^{\rm VV'}
    ,
    \\
    & F_{ab}^{\rm VV'}
    \coloneqq
    \sum_{\lambda} 
    p(\lambda)
    F_{a|\lambda}^{\rm V}
    \ot
    F_{b|\lambda}^{\rm V'}, 
    \\
    &F_{a|\lambda}^V
    \coloneqq 
    \tr_{A}
    \left[
    N_a^{VA}
    \left(
    \mathds{1}^V
    \otimes
    \rho_\lambda^{A}
    \right)
    \right],
    \\
    &F_{b|\lambda}^{V'}
    \coloneqq 
    \tr_{B}
    \left[
    N_b^{BV'}
    \left(
    \rho_\lambda^{B}
    \otimes
    \mathds{1}^{V'}
    \right)
    \right],
    \end{align}} 
    with $\mathbb{N}^{VA}$ $\mathbb{N}^{BV}$ bipartite POVMs, $\{\rho_\lambda^A\}$, $\{\rho_\lambda^B\}$ sets of states, $p_\Lambda$ a PMF, and $\{L_{ab}^{\rm VV'} \}$ a distributed POVM.
\end{lemma}
We label the cone of all free Buscemi operators $\big\{F_{ab}^{\rm VV'}\big\}$ with $\fbn$ and define the dual cone $\fbn^*$. We also say that a set of Buscemi operators $\big\{F_{ab}^{\rm VV'}\big\} \in \fbn$ if and only if there is a PMF $p_\Lambda$, bipartite POVMs $\mathbb{N}^{VA}$, $\mathbb{N}^{BV'}$, sets of states $\{\rho_\lambda^A\}$, $\{\rho_\lambda^B\}$, such that we have $F_{ab}^{\rm VV'} = \sum_{\lambda} p(\lambda) F_{a|\lambda}^{\rm V} \ot F_{b|\lambda}^{\rm V'}$, with $F_{a|\lambda}^V
=
\tr_{A}
[
    N_a^{VA}
    (
        \mathds{1}^V
        \otimes
        \rho_\lambda^{A}
    )
]
$,
and
$
F_{b|\lambda}^{V'}
=
\tr_{B}
[
    N_b^{BV'}
    (
        \rho_\lambda^{B}
        \otimes
        \mathds{1}^{V'}
    )
]
$, $\forall a$. This definition is equivalent to the one given in \eqref{eq:RRoBN} under the assumption that input sets of states $\{\omega_x^{\rm V}\}$, $\{\omega_y^{\rm V'}\}$ are tomographically-complete. In other words, $\{F_{ab}^{\rm VV'}\} \in \fbn$ if and only if any behaviour $\{p(ab|\omega_x\omega_y)\}$ arising from the measurement specified by $\{F_{ab}^{\rm VV'}\}$ is free according to \eqref{eq:RRoBN}. Any element of the dual cone $\{W_{ab}^{\rm VV'}\} \in \fbn^*$ satisfies:
\begin{align}
    \sum_{a,b}
    \tr
    \left[
    W_{ab}^{\rm V'V}
    F_{ab}^{\rm V'V}
    \right] 
    \geq
    0,
    \qquad
    \forall \, \{F_{ab}^{\rm V'V}\} \in \fbn.
\end{align}
Let us now write the dual form of the optimisation problem from \cref{robn-primal}, i.e:
\begin{lemma}[RoBN - dual] 
    Let $\{\omega_x^{\rm V}\}$ and $\{\omega_y^{\rm V'}\}$ be tomographically-complete sets of states. Then the dual of RoBN can be written as: 
    \begin{align}
        R_\mathrm{BN}
        &(
        \{M_a^{\rm VA}\}, 
        \{M_b^{\rm BV'}\}, 
        \rho^{\rm AB}
        ) = 
        \label{eq:robnd0}
        \\ 
        &\qquad \qquad 
        \max_{
        W_{ab}^{\rm VV'}
        } \,\, 
        -\sum_{a, b}
        \tr
        \left[
        M_{ab}^{\rm VV'}
        W_{ab}^{\rm VV'}
        \right]
        ,
        \nonumber
    \end{align}
    subject to:
    \begin{align}
        & \{W_{ab}^{\rm VV'}\} 
        \in
        \fbn^* 
        ,
        \label{eq:robnd1}
        \\
        & \sum_{a,b}
        \tr
        \left[
        W_{ab}^{\rm VV'}
        \right]
        \leq
        o_{\rm A}
        \cdot
        o_{\rm B}.
        \label{eq:robnd2}
    \end{align}
\end{lemma}

\section{Proof of \cref{r:r2}}
\label{a:a2}

Let us begin by showing that:
\begin{align}
    \label{eq_pr2_8}
    \max_{
    \{M_{b}^{\rm BV'}\}
    }\,\, 
    R_\mathrm{BN}
    (
    \{M_{a}^{\rm VA}\}, 
    &\{M_{b}^{\rm BV'}\}, 
    \rho^{\rm AB}
    ) 
    \leq
    \\ 
    & R_\mathrm{T}
    (
    \{M_a^{\rm VA}\},
    \rho^{\rm AB}
    )
    .
    \nonumber
\end{align}

\begin{proof}(of \eqref{eq_pr2_8})
Fix a POVM $\{M_{a}^{\rm VA}\}$, a quantum state $\rho^{\rm AB}$, and let us assume we have solved the primal conic program for $R_\mathrm{T}(\{M_a^{\rm VA}\}, \rho^{\rm AB})$  using variables: $r$, $\{p(\lambda)\}$, $\{N_{a|\lambda}^{\rm V'}\}$, $\{\sigma_{\lambda}^{\rm B}\}$, $\{K_a^{\rm V'B}\}$ (see \cref{rot_choi}). These primal variables for all $a$ satisfy:
\begin{align}
\label{eq_pr2_5}
    J_a^{\rm V'B} 
    +
    r
    K_a^{\rm V'B}
    = 
    (1+r)
    \sum_{x,\lambda}
    p({\lambda})\,
    N_{a|\lambda}^{\rm V'}
    \ot
    \sigma_{\lambda}^{\rm B}.
\end{align}
Consider now an arbitrary POVM on Bob's side $\{M_b^{\rm B\widetilde{V}}\}$, and let us transform the above operator equality \eqref{eq_pr2_5} by applying a linear map $(\cdot) \rightarrow \tr_{\rm B} [((\cdot) \ot \mathds{1}^{\rm \widetilde{V}})(\mathds{1}^{\rm V'} \ot M_{b}^{\rm B\widetilde{V}}) ]$, we can write: 
\begin{align}
    \nonumber
    J_a^{\rm V'B} 
    &\rightarrow
    \tr_{\rm B}
    \left[
    (J_a^{\rm V'B}
    \ot
    \mathds{1}^{\rm \widetilde{V}})
    (
    \mathds{1}^{\rm V'}
    \ot 
    M_{b}^{\rm B\widetilde{V}}
    )
    \right] \\ 
    &
    =
    (
        M_{ab}^{\rm V'\widetilde{V}}
    )^{T_{\rm V'}},
    \\ 
    \nonumber
    K_a^{V'B} 
    &\rightarrow
    \tr_{\rm B}
    \left[
    \left(
        K_a^{V'B}
        \ot
        \mathds{1}^{\rm \widetilde{V}}
    \right)
    (
    \mathds{1}^{\rm V'} 
    \ot
    M_{b}^{\rm B\widetilde{V}}
    )
    \right]
    \\
    &
    =
    (
    N_{ab}^{V'\widetilde{V}}
    )^{T_{V'}}
    \\
    N_{a|\lambda}^{\rm V'}
    \ot
    \sigma_{\lambda}^{\rm B} 
    &\rightarrow
    \tr_{\rm B}
    \left[
    ( 
    N_{a|\lambda}^{\rm V'}
    \ot
    \sigma_{\lambda}^{\rm B} 
    \ot
    \mathds{1}^{\rm \widetilde{V}}
    )
    (
    \mathds{1}^{\rm V'} 
    \ot
    M_{b}^{\rm B\widetilde{V}}
    )
    \right] \nonumber \\
    &= 
    N_{a|\lambda}^{\rm V'}
    \ot
    N_{b|\lambda}^{\rm \widetilde{V}}
    .
\end{align}
In the first line $\{M_{ab}^{\rm V'\widetilde{V}}\}$ is the distributed POVM associated to the triple ($\rho^{AB}$, $\{M_{a}^{\rm VA}\}$, $\{M_{b}^{\rm B\widetilde{V}}\}$), and use that for any bipartite linear operator $O^{V\widetilde{V}}$ we have (see also Appendix A of Ref \cite{NCT1}):
\begin{align}
    \tr_V 
    [
    (
    \phi_{+}^{V'V}
    \ot
    \mathds{1}^{\widetilde{V}}
    )
    (
    \mathds{1}^{V'}
    \ot
    O^{V\widetilde{V}}
    )
    ]
    =
    \frac{1}{d}
    (O^{V'\widetilde{V}})^{T_{V'}}
    ,
    \label{eq:si}
\end{align}
with $T_{\rm V'}$ the partial transpose with respect to system $V'$. In the second line $\{N_{ab}^{\rm V'\widetilde{V}}\}$ is the distributed POVM associated to the triple ($\sigma^{AB}$, $\{N_{a}^{\rm VA}\}$, $\{M_{b}^{\rm B\widetilde{V}}\}$) and similarly follows from \eqref{eq:si}. In the third line we use
$
N_{b|\lambda}^{\rm \widetilde{V}} 
\coloneqq 
\tr_{\rm B}
[
M_b^{\rm B\widetilde{V}}
(
\sigma_{\lambda}^{\rm B} 
\ot
\mathds{1}^{\rm \widetilde{V}}
)
]
$. We can now infer that (\ref{eq_pr2_5}) implies:
\begin{align}
    \label{eq_pr2_6}
    (
    M_{ab}^{\rm V'\widetilde{V}}
    )^{T_{\rm V'}}\!\! 
    +
    r
    (
    N_{ab}^{V'\widetilde{V}}
    )^{T_{\rm V'}}\!\!
    =\
    &
    (1+r)
    \sum_{\lambda} 
    p(\lambda)
    N_{a|\lambda}^{\rm V'}
    \ot
    N_{b|\lambda}^{\rm\widetilde{V}}
    .
\end{align}
Let us now relabel subsystems $\rm V'\widetilde{V} \rightarrow \rm VV'$. Taking partial transpose with respect to the first system ($\rm V$) of \eqref{eq_pr2_6} leads to:
\begin{align}
    \label{eq_pr2_7}
    M_{ab}^{\rm VV'} 
    +
    r
    N_{ab}^{VV'} 
    =
    \sum_{\lambda}
    p(\lambda)
    (N_{a|\lambda}^{\rm V})^T
    \ot
    N_{b|\lambda}^{\rm V'}.
\end{align}
With this let us now construct a sub-optimal guess for the primal variables appearing in primal conic program of $R_\mathrm{BN}(\{M_{a}^{\rm VA}\}, \{M_{b}^{\rm BV'}\}, \rho^{\rm AB})$ (see \cref{robn-primal}) as:
{\small\begin{align}
  r 
  &=
  R_\mathrm{T}
  (
  \{M_a^{\rm VA}\},
  \rho^{\rm AB}
  ),
  \\
  F_{a|\lambda}^{\rm V} 
  &=
  (N_{a|\lambda}^{\rm V})^T
  , 
  \\
  p^*(\lambda) 
  &=
  p(\lambda),
  \\  
  F_{b|\lambda}^{\rm V'} 
  &=
  N_{b|\lambda}^{\rm V'} 
  =
  \tr_{\rm B}[
  M_b^{\rm B\widetilde{V}}
  (
  \sigma_{\lambda}^{\rm B}
  \ot
  \mathds{1}^{\rm \widetilde{V}}
  )], 
  \\
  L_{ab}^{VV'}
  &=
  N_{ab}^{VV'}
  \hspace{-0.2cm}
  =
  (
  \tr_{\rm B}
    [
    (
        K_a^{VB} 
        \hspace{-0.1cm}
        \ot
        \mathds{1}^{V'}
    )
    (
        \mathds{1}^{\rm V} 
        \hspace{-0.1cm}
        \ot
        M_{b}^{\rm BV'}
    )
    ]
    )^{T_{V'}}
    .
\end{align}}
By Eq. (\ref{eq_pr2_7}) we can infer that this is a feasible set of primal variables for RoBN. This then shows that the sub-optimal guess of variables (C8-C11) leads to RoBN no larger than RoT and, the primal conic program of RoBN being a minimisation then implies:
{\small \begin{align}
    R_\mathrm{BN}
    (
    \{M_{a}^{\rm VA}\}, 
    &\{M_{b}^{\rm BV'}\}, 
    \rho^{\rm AB}
    ) 
    \leq
    R_\mathrm{T}
    (
    \{M_a^{\rm VA}\},
    \rho^{\rm AB}
    )
    .
    \label{eq:lastt}
\end{align}}
This construction holds for any POVM on Bob's side $\{M_b^{\rm BV'}\}$, and so it holds, in particular, for the POVM achieving the maximisation in the LHS model of \eqref{eq:lastt}, and therefore proving inequality \eqref{eq_pr2_8}. 
\end{proof}

Let us now prove the reverse inequality, that is:
\begin{align}
    \label{eq_pr2_1}
    \max_{
    \{M_{b}^{\rm BV'}\}
    }
    \,\, 
    R_\mathrm{BN}
    (
    \{M_{a}^{\rm VA}\}, 
    &\{M_{b}^{\rm BV'}\}, 
    \rho^{\rm AB}
    )
    \geq \\
    \nonumber
    &R_\mathrm{T}
    (
    \{M_{a}^{\rm VA}\},
    \rho^{\rm AB}
    ).
\end{align}
\begin{proof} (of \eqref{eq_pr2_1})
Fix a POVM
$
\{M_{a}^{\rm VA}\}$, a quantum state $\rho^{\rm AB}
$, and let us assume we have solved the dual problem for $R_\mathrm{T}(\{M_{a}^{\rm VA}\},\rho^{\rm AB})$  from \cref{rot_choi_dual}. Let us use from this solution the teleportation witness $\{W_a^{\rm V'B}\} \in \ftel^*$ satisfying:
\begin{align}
    R_\mathrm{T}
    (\{M_a^{\rm VA}\},
    \rho^{\rm AB})
    =
    -\sum_{a}
    \tr
    \left[
        J_a^{\rm V'B}
        W_{a}^{\rm V'B}
    \right].
    \label{eq:previous1}
\end{align}
Let $\{U_b^{\rm V'}\}$ for $b \in \{1, \ldots, o_{\rm B}\}$, $o_{\rm B} = d^2$ be a set of generalised Pauli operators with respect to a basis $\{\ket{i}^{\rm V'}\}$. Consider now a potentially suboptimal POVM $\{M_{b}^{\rm BV'}\}$ in the LHS model of \eqref{eq_pr2_1} as well as a potentially suboptimal dual witness \{$W_{ab}^{\rm V'V} \} \in \fbn^*$ (as per \eqref{eq:robnd0}) as:
\begin{align}
     &M_{b}^{\rm BV'} 
     \coloneqq
     (
     \mathcal{I} 
     \ot
     \mathcal{U}_{\rm b}^{\rm V'}
     )
     [
     \phi_+^{\rm BV'}
     ],
     \label{eq:guess1}
     \\
     &W_{ab}^{\rm VV'} 
     \coloneqq
     \frac{1}{d}
     (
     \mathcal{I}
     \ot
     (\mathcal{U}_{\rm b}^{\rm V'})^{\dagger}
     )
     [
     (
     W_a^{\rm V V'}
     )^{T}
     ]
     ,
     \label{eq:guess2}
\end{align}
with $\mathcal{U}_{b}^{\rm V'}[\cdot] \coloneqq U_{b}^{\rm V'}(\cdot)(U_{b}^{\rm V'})^{\dagger}$ a unitary channel and $W_a^{\rm V V'}$ being the same operator as $W_a^{\rm V'B}$ but acting on Hilbert space associated with $\rm{VV'}$. Before proceeding to show that these operators constitute a  ``good" guess, we first need to check that they form a ``valid" guess. It is straightforward to check that $\{M_{b}^{\rm BV'}\}$ is a POVM, and so we next check that $\{W_{ab}^{\rm VV'}\}$ is a ``valid" witness for Buscemi nonlocality. Explicitly, this means checking the conditions \eqref{eq:robnd1} and \eqref{eq:robnd2}. In order to show that $\{W_{ab}^{\rm VV'}\} \in \fbn^*$ \eqref{eq:robnd1}, notice that $\{W_a^{\rm V'B}\} \in \ftel^*$ and therefore for all PMFs $p_\Lambda$, all POVMs $\{N_{a|\lambda}^{\rm V'}\}$ and all sets of states $\{\sigma_{\lambda}^{\rm B}\}$ the following holds:
\begin{align}
\label{eq_pr2_3}
     &\sum_{a, \lambda}
     p(\lambda)
     \tr
     [
     W_a^{\rm V'B}
     (
     N_{a|\lambda}^{\rm V'}
     \ot
     \sigma_{\lambda}^{\rm B}
     )
     ]
     \geq
     0, \\
     \label{eq_pr2_4}
     &\sum_a 
     \tr 
     [
     W_a^{\rm V'B}
     ]
     \leq
     o_{\rm A}
     \cdot
     d.
\end{align}
With this, let us consider an arbitrary free generalised measurement $\{F_{ab}^{\rm VV'}\} \in \fbn$. We can write:
\begin{align}
    &\sum_{a,b}  
    \tr
    \left[
    W_{ab}^{\rm VV'}
    F_{ab}^{VV'} 
    \right]
    \nonumber
    \\
    &= 
    \sum_{a,b,\lambda}
    p(\lambda)
    \tr
    \left[
    \frac{1}{d}
    (
    \mathcal{I}
    \ot
    \mathcal{U}^{\dagger}_{\rm b}
    )
    [
    W_a^{T}
    ]
    \cdot
    F_{a|\lambda}^{\rm V}
    \ot
    F_{b|\lambda}^{\rm V'}
    \right] \nonumber\\
    &= \frac{1}{d} 
    \sum_{a, b, \lambda}
    p(\lambda)
    \tr
    \left[
    W_a^{T}
    \cdot
    F_{a|\lambda}^{\rm V}
    \ot
    \mathcal{U}_{b} 
    \left[
    F_{b|\lambda}^{\rm V'}
    \right]
    \right] \nonumber\\
    &= \frac{1}{d}
    \sum_{a,b,\lambda}
    p(\lambda)
    \tr
    \left[
    W_a^{T} 
    \cdot
    F_{a|\lambda}^{\rm V}
    \ot
    \mathcal{U}_{b}
    \left[
    F_{b|\lambda}^{\rm V'}
    \right] 
    \right] \nonumber\\
    &= 
    \sum_{a, \lambda}
    p(\lambda)
    \tr
    \left[
    W_a^T
    \cdot
    F_{a|\lambda}^{\rm V} 
    \ot
    \sigma_{\lambda}^{\rm V'}
    \right],
    \label{eq_pr2_2}
\end{align}
where in the last line we have introduced:
\begin{align}
   \sigma_{\lambda}^{\rm V'} 
   \coloneqq
   \frac{
    \sum_b
   \mathcal{U}_{b}
   \left[
   F_{b|\lambda}^{\rm V'}
   \right]
   }{
   \sum_b
   \tr
   \mathcal{U}_{b}
   \left[
   F_{b|\lambda}^{\rm V'}\right]}
   =
   \frac{1}{d}
   \sum_b
   \mathcal{U}_{b}
   \left[
   F_{b|\lambda}^{\rm V'}
   \right].
\end{align}
Denoting $\widetilde{N}_{a|\lambda} \coloneqq (F_{a|\lambda}^{\rm V})^T$ and $\widetilde{\sigma}_{\lambda} \coloneqq (\sigma_{\lambda}^{V'})^T$ we can continue and write \eqref{eq_pr2_2} as:
{\small \begin{align}
    \sum_{a,b} 
    \tr
    \left[
        W_{ab}^{\rm VV'}
        F_{ab}^{VV'} 
    \right]
    &= 
    \sum_{a, \lambda} 
    p(\lambda) 
    \tr
    \left[
        W_a(N_{a|\lambda} 
        \ot
        \sigma_{\lambda})
    \right]
    \\
    &\geq 
    0,
\end{align}}
which follows from the fact that $\{W_a\}$ is a teleportation witness, i.e. \eqref{eq_pr2_3} holds. This then shows that $\{W_{ab}^{\rm VV'}\} \in \fbn^*$ \eqref{eq:robnd1}. It remains to show that $\{W^{\rm VV'}_{ab}\}$ is bounded  \eqref{eq:robnd2}. We can see this as follows:
\begin{align}
    \sum_{a,b} 
    \tr
    W_{ab}^{\rm VV'} 
    &=
    \frac{1}{d}
    \sum_{a,b}
    \tr
    [
        (
            \mathcal{I}
            \ot
            \mathcal{U}_{b}^{\dagger}
        )
        [
        W_a^T
        ]
    ],
    \nonumber\\
    &= 
    \frac{1}{d}
    \sum_{a,b}
    \tr
    \left[W_a^T\right],
    \nonumber\\
    &= 
    o_{\rm B} 
    \cdot
    \frac{1}{d} 
    \sum_{a}
    \tr 
    \left[W_a\right],
    \nonumber\\
    &\leq 
    o_{\rm A}
    \,
    o_{\rm B}.
\end{align}
In the first line we replace \eqref{eq:guess2}. In the second line we use that the trace is invariant under unitaries and transposition. In the third line we sum over $b$. In the fourth line we use that $\{W_a\}$ is a teleportation witness, see \eqref{eq_pr2_4}. This then concludes the proof that the guess $\{W_{ab}^{\rm VV'}\}$ \eqref{eq:guess2} is a valid witness for Buscemi nonlocality. Let us now return to our primary goal, that is showing the inequality from \eqref{eq_pr2_1}. Using our choices for $\{M_{b}^{\rm BV'}\}$ \eqref{eq:guess1} and $\{W_{ab}^{\rm VV'}\}$ \eqref{eq:guess2}, we get:
\begin{align}
    &\max_{\{M_{b}^{\rm BV'}\}}\,\, 
    R_\mathrm{BN}
    (
        \{M_{a}^{\rm VA}\}, 
        \{M_{b}^{\rm BV'}\}, 
        \rho^{\rm AB}
    )  
    \nonumber
    \\
    &\overset{1}{\geq} 
    - \sum_{a,b} 
    \tr
    \left[
        W_{ab}^{\rm VV'}
        M_{ab}^{\rm VV'}
    \right] \nonumber\\
    &\overset{2}{=} 
    - \frac{1}{d^2}
    \sum_{a,b}
    \tr
    \left[
        W_a^{\rm BV'}
        J_{a}^{\rm BV'}
    \right] \nonumber\\
    &\overset{3}{=} 
    \frac{o_{\rm B}}{d^2} \,
    R_\mathrm{T}
    \left(
        \{M_a^{\rm VA}\},
        \rho^{\rm AB}
    \right) \nonumber\\
    &
    \overset{4}{=} 
    R_\mathrm{T}
    \left(
        \{M_a^{\rm VA}\}, 
        \rho^{\rm AB}
    \right).
\end{align}
In the first line we start from the dual of RoBN \eqref{eq:robnd0}, and use that we are using a potentially suboptimal guess $\{W_{ab}^{\rm VV'}\}$. In the second line we replace $\{W_{ab}^{\rm VV'}\}$ \eqref{eq:guess2}, and replace $\{M_{b}^{\rm BV'}\}$ \eqref{eq:guess1} in $\{M_{ab}^{\rm VV'}\}$ \eqref{eq:buscemimeas}. The step is completed by noting that by defining
\begin{align}
        \!\! 
        \tilde{J_a}^{\rm VV'}
        & \! 
        \coloneqq
        \!\tr_{\rm AB}
        \left[
            \left(
                M_a^{\rm VA} \ot \phi_+^{\rm{V'B}}
            \right)
            \left(
                 \mathds{1}^{\rm VV'}
                \!\!\ot\! 
                \rho^{\rm AB}
            \right)
        \right],
\end{align}
we obtain the relationship $\tr[\tilde{J_a}^{\rm VV'} W_a^{VV'}] = {\frac{1}{d}}\tr[J_a^{\rm V'B} W_a^{V'B}]$, $\forall a$, with $J_a^{\rm V'B}$ as in \eqref{eq:CJ}. This latter relationship follows from the more general identity:
\begin{align}
    &\tr
    \left[
        \left(
            M^{\rm VA} 
            \ot 
            \phi_+^{\rm{BV'}}
        \right)
        \left(
            (W^{\rm VV'})^T
            \!\!\ot\! 
            P^{\rm AB}
        \right)
    \right]
    \nonumber
    \\
    &=
    {\frac{1}{d}}
    \tr
    \left[
        \left(
            M^{\rm VA} 
            \ot 
            W^{\rm BV'}
        \right)
        \left(
             \phi_+^{\rm{VV'}}
            \!\!\ot\! 
            P^{\rm AB}
        \right)
    \right]
    ,
\end{align}
which holds true for all linear bipartite operators $M^{VA}, P^{AB}$, $W^{VV'}(W^{BV'})$,  and which can be proven in a similar manner as \eqref{eq:transp} and \eqref{eq:si}, or by following a diagrammatic approach as in \cite{DA1, DA2, DA3}. This, on the other hand, implies
\begin{align}
    &\tr[W_{ab}^{VV'} M_{ab}^{VV'}]
    \nonumber\\
    &= {\frac{1}{d}}
    \tr[W_a^{VV'} 
    (U_b^{\dagger} 
    \ot 
    \mathcal{I})
    (U_b \ot \mathcal{I}) 
    \tilde{J_a}^{VV'}] \nonumber\\
    &= {\frac{1}{d}}
    \tr[W_a^{VV'} 
    \tilde{J_a}^{VV'}] \nonumber\\
    &= 
    \frac{1}{d^2} 
    \tr[W_a^{V'B} J_a^{V'B}].
\end{align}
In the third line we sum over $b$ and use \eqref{eq:previous1}. In the fourth line we used that $o_{\rm B} = d^2$. This concludes the proof of \eqref{eq_pr2_1}. Finally, inequality \eqref{eq_pr2_8}  along with \eqref{eq_pr2_1} completes the proof of the full statement in \cref{r:r2}.  
\end{proof}

\end{document}